\documentclass[10pt]{llncs} 

\DeclareFontShape{OT1}{cmtt}{bx}{n}{<5><6><7><8><9><10><10.95><12><14.4><17.28><20.74><24.88>cmttb10}{}

\usepackage{graphicx}
\usepackage{amssymb}
\usepackage{lineno}
\usepackage{verbatim}
\usepackage{listings}
\usepackage{todonotes}
\usepackage{cite}
\usepackage{zed-csp}

\begin{document}

\title{Microservices:\\ How To Make Your Application Scale}
\author{Nicola Dragoni\inst{1,5}, Ivan Lanese\inst{2},  Stephan Thordal Larsen \inst{1} \\ Manuel Mazzara\inst{3}, Ruslan Mustafin\inst{3}, Larisa Safina\inst{3,4}}

\institute{
Technical University of Denmark\\
\email{ndra@dtu.dk, stephan@thordal.io}
\and
University of Bologna/INRIA \\
\email{ivan.lanese@gmail.com}
\and
Innopolis University, Russian Federation\\
\email{\{l.safina, m.mazzara, r.mustafin\}@innopolis.ru}
\and
University of Southern Denmark\\
\and
\"{O}rebro University, Sweden \\
}

\maketitle

\begin{abstract}
The microservice architecture is a style inspired by service-oriented computing that has recently started gaining popularity and that promises to change the way in which software is perceived, conceived and designed. In this paper, we describe the main features of microservices and highlight how these features improve scalability.
\end{abstract}

\section{Introduction}
History of programming languages and paradigms has been characterized in the last few decades by a progressive shift towards distribution, modularization and loose coupling, with the purpose of increasing code reuse and robustness \cite{AlmeidaALGM04}. This necessity has been dictated by the need of increasing software quality, not only in safety and financial-critical applications, but also in more common off-the-shelf software packages. 

Service oriented architectures (SOAs) can be seen as a step in this direction, where the need for code reuse and robustness was coupled with the need for interoperability between heterogeneous information systems, possibly belonging to different companies. This brought up the idea of a \emph{service} as a software entity interacting with other software entities via message passing communications using standard data formats and protocols (e.g., XML, SOAP and HTTP) and well-defined interfaces.

Microservices are a further step along this road, emphasizing the use of small services, called indeed microservices, and moving the service oriented techniques from system integration to system design, development and deployment.

The microservice architecture \cite{ms-pause} is built on a few basic principles:

\begin{itemize}
	\item \textit{Bounded Context}. First introduced in \cite{evans2004domain}, bounded context means that related functionalities are combined into a single business capability, and each microservice implements one such capability. In this way there is a perfect alignment between business capabilities and system structure, making it easy, e.g., to know where a functionality is, in order to update or fix it.    
	\item \textit{Size}. The focus on small size is a crucial novelty of microservices w.r.t.\ the previous SOAs.
 Idiomatic use of microservice architectures suggests that if a service is too large, it should be refined into two or more services, thus preserving granularity and maintaining focus on providing a single business capability only. The small size brings major benefits in terms of service maintainability and extendability: a small service can be easily modified, and if needed rebuilt from scratch with limited resources and in limited time. 
	\item \textit{Independency}. This concept encourages loose coupling and high cohesion by stating that each microservice in microservice architectures is operationally independent from others, and the only form of communication between services is through their published interfaces. This is fundamental since this allows one to change, fix or upgrade a microservice without compromising the system correctness, provided that the interfaces are preserved. 
\end{itemize}

The shift towards microservices is a sensitive matter these days. Several companies are involved in a major refactoring of their back-end systems to accommodate the advantages of the new paradigm. Other companies just start their business model developing software following the microservice paradigm since day one. We are in the middle of a major change in the view in which software is intended, and in the way in which capabilities are organized into components, and industrial systems are conceived. In the next section we highlight another advantage of microservices: scalability, for performance, fault tolerance or availability reasons.

\section{Scalability}
Scalability is one of the key features provided by the microservice paradigm. In this section, we aim at giving an overview on how microservice characteristics naturally contribute to system scalability. We emphasize that while frequently scalability is needed for performance reasons, to cope with high load, scalability can also be used to ensure availability and fault tolerance. According to the reason why scalability is needed, slightly different approaches need to be used, as we will emphasize below.\\

\noindent \textbf{Distribution.} Distribution is not an original feature of microservices, since, e.g., SOAs are distributed as well. However, thanks to their small size, microservices take this characteristics to an extreme: each business capability, including their functionalities and the related data, is realized by an independent service, which can be deployed on a host possibly different from the one of other microservices of the same application. As first result, this causes a natural distribution of the workload that can make the system significantly more efficient than a monolith \cite{Bondi2000}. Distribution also makes microservice architectures highly available, since the failure of a single microservice does not necessarily result in the failure of other microservices. Distribution can also utilize locality and locate services closer to the clients they serve, resulting in better geographical scalability \cite{Bondi2000,Neuman94}.\\

\noindent \textbf{Non-uniform scaling.} Typically, when monolithic architectures are exposed to growing load, it is difficult to locate which components of the system are actually affected, since the system runs within a single process. This means that, although only a single component may be experiencing load, the whole monolith will need to scale, e.g. by replication or vertical scaling. Even if it is known which is the component that is experiencing load, it is difficult to scale it in isolation. The same reasoning may apply to SOAs: services in SOAs may be large, frequently hiding a whole monolithic application behind a service-oriented interface, hence they may only scale at a large granularity. 
The same applies when scalability is needed to implement high availability: if only some components of a monolith or of a large service are required to be highly available, the whole monolith/large service will have to be highly available. 

Since microservices are implemented and deployed independently of each other, i.e. they run within independent processes, they can be monitored and scaled independently, as shown by the example below.

\begin{figure}[!ht]
\centering
\includegraphics[width=0.5\textwidth]{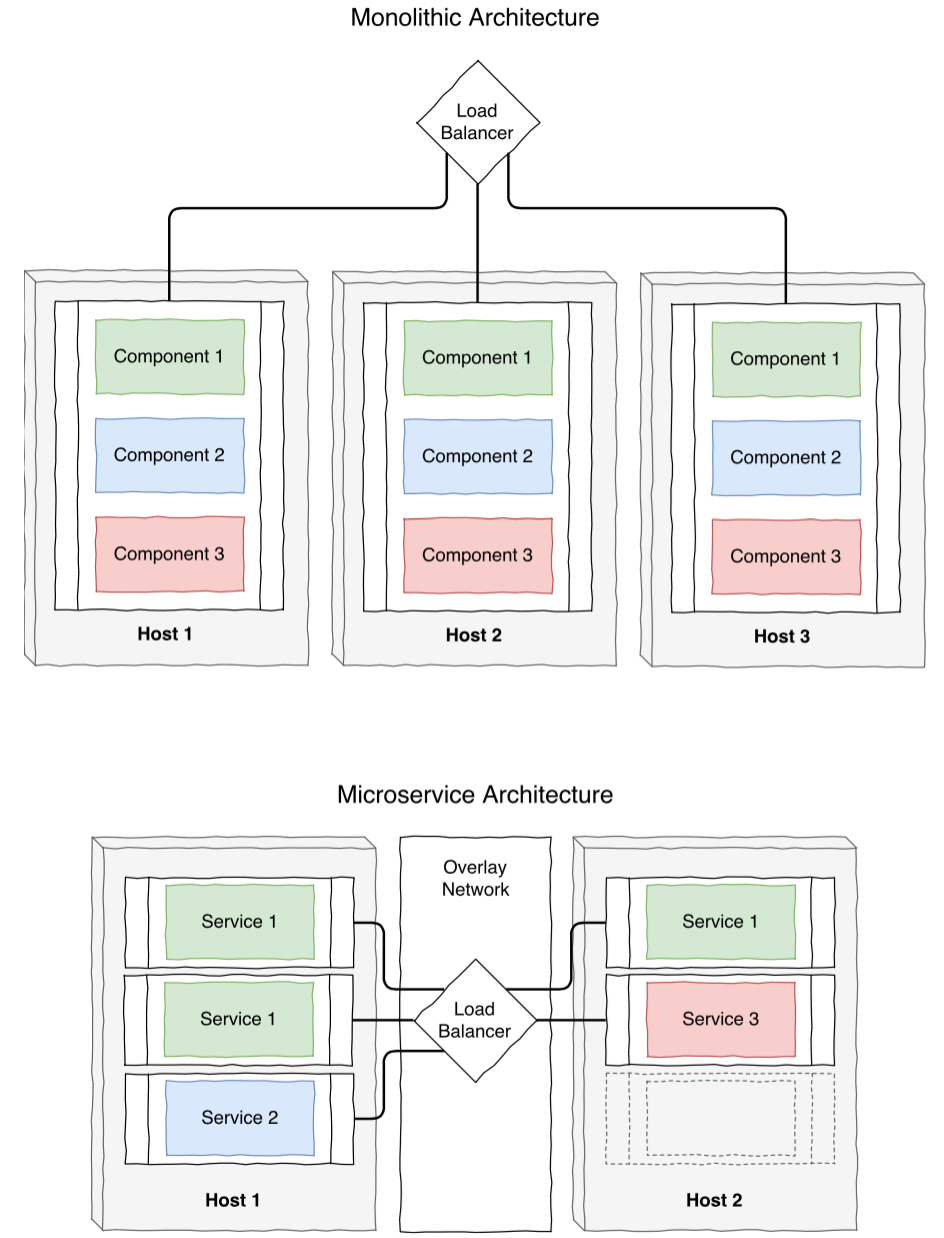}
\caption{Scaling in microservices vs monolithic architecture \label{comparison}}
\end{figure}

\noindent \emph{Example.} A simplified illustration showcasing the benefits of scaling a microservice architecture, compared to a monolithic architecture, is given in Figure \ref{comparison}. Both of the two systems implement componentization of software, the monolith utilizing regular software components, such as libraries, and the microservice architecture utilizing microservices, i.e. \emph{Component x} corresponds to \emph{Service x}. In this scenario \emph{Component/Service 1} is experiencing a load that requires to replicate it to 3 instances. Since the monolith is deployed as a single process, one needs to replicate the whole system, including all 3 components, across three hosts. In a microservice architecture one can simply replicate the single service experiencing load, resulting in allocation of much fewer hosts. The load balancers are in place in both systems to split the load across replicas. However, in the monolith the balancer splits only external requests, while in the case of the microservice architecture it splits both external requests and internal requests between the different microservices, thus allowing for a more uniform load balancing. This happens in particular when external requests may trigger computations which are heavy in a possibly non-uniform way: only balancing external requests may not be enough.

%

The reliance on \emph{Domain-Driven Design} \cite{evans2004domain} and the strive towards \emph{high cohesion} means that growing load will typically be delimited to a subset of associated microservices \cite{newman2015building}. The specific microservices actually experiencing the growing load can then be scaled, e.g., by relocating them to the more performant hosts or by replicating them across a cluster or on the cloud.

A similar argument applies to the technology adopted for implementing each microservice: the technology used to build a microservice can be chosen in order for it to perform at best. For instance, a computation-intensive microservice might be implemented in C++, while a microservice requiring to handle complex types could be implemented in a language with a sophisticated type system, like Haskell. This is not possible in a monolithic architecture which is typically bound to a single platform and language.\\

\noindent \textbf{Portability.} Microservices are typically packaged in \emph{containers}, as provided, e.g., by \emph{Docker}~\cite{docker} or similar technologies. A container includes the microservice and all its environment (libraries, databases, \dots) in a unique entity which can be easily deployed on any platform supporting the chosen container technology, ensuring uniform behavior over heterogeneous platforms (hosts, data-centers and cloud providers) and isolation w.r.t. other containers (e.g., different microservices can use different versions of the same library without conflicts).
The portability ensured by containers enables effortless relocation or replication of a microservice across heterogeneous platforms. Microservice architectures are therefore ideal for scaling a system horizontally, since the microservices can easily be relocated to newly provisioned hosts. \\

\noindent \textbf{Elasticity.} The ability to easily replicate individual microservices, coupled with the ability to locate both a single and multiple microservices on a single host, also enables microservice architectures to be elastic, that is to dynamically scale according to the load. Because of this multiple-service-per-host model, deploying a microservice architecture to a dynamically-sized cluster, and in particular on the Cloud, allows it to utilize available resources very efficiently. When the load is high, the system can easily be expanded by exploiting additional hosts dynamically allocated to it in the cluster or new virtual machines on the cloud, and when resources become redundant because of lower load then those hosts can be de-provisioned and removed from the cluster/cloud again. In the same way, the number of service replicas can be increased or shrunk when needed. This feature makes microservices a natural technology for the cloud, and suggests that microservice popularity will continue to grow as far as more and more applications are moved to the cloud.\\

\noindent \textbf{Availability} We have already highlighted some of the ways in which microservices can help availability, but here we  summarize the main aspects related to the topic. In general, high availability is achieved by microservices' ability to be replicated and spread across data-centers and geographical distances, allowing them to spread load and cope with failing and congested hardware.
Another relevant aspect concerns system update and evolution: while updating a monolithic application requires to stop it and re-deploy it, thus causing a possibly long downtime, replicability and independence allow microservices to solve the problem. First, updating a microservice architecture normally involves just one or a few microservices related to the business capability that needs to be fixed or improved, hence reducing the deployment time. Furthermore, the old and the new version of the same microservice can run side by side, e.g., the old one completing running requests and the new one taking care of new requests. The old one can then be removed when its job is ended. Note that containerization avoids interferences between the two versions of the service, e.g., allowing them to rely on different versions of the same library. This naturally leads to smaller but more frequent updates, in the direction of continuous deployment.\\

\noindent \textbf{Robustness.} As for availability, also robustness benefits from using a microservice approach. Indeed, one may replicate microservices as described above to ensure fault tolerance. Fault tolerance however is also naturally improved because of the usage of \emph{containerization} and independent processes. Indeed, a single microservice is completely isolated from other microservices and can only be affected by them through its defined interfaces or through the resources it relies on. This means that even though some microservices might fail, isolation ensures that other microservices and their environments are not affected. Of course, this requires microservices to implement some fault-tolerant mechanisms that can detect possible failures in microservices they depend on in order to prevent cascading failures.

One should however pay attention that low level interferences may still happen, in particular when multiple microservices are deployed on the same host. Indeed, although their logical environments might be isolated, their physical one is not. If a single microservice consumes all the resources of a host shared with other microservices, those microservices will be affected. Therefore one should be careful when placing microservices together on the same host and take the possible load of each of the microservices into account both before and during operation, ensuring that resources are not exhausted by a single microservice. \\


\noindent \textbf{No silver bullet.} The description above should clarify how microservices provide a natural way to reach scalability, including availability and fault tolerance. However, this does not come for free: having multiple independent entities introduce some extra administrative overhead, in particular for deployment, administration, monitoring, and security. While there are approaches to mitigate these problems (but still far from satisfactory, at least concerning security), this means that sometimes microservices are not the solution. The description above should help to understand the advantages of microservices, and deciding whether they are a good technique for the problem at hand. We discuss below some concrete cases to further clarify the issue.

\section{The Language Choice}
While microservice architectures can be built using a wide range of technologies, possibly combined into the same system, we think that the use of a dedicated language can simplify the development of microservice systems. Our experience is based on the language Jolie, the only language we are aware of natively supporting microservice architectures. While we refer to~\cite{MGZ14} for a detailed description of the Jolie language, we recall here its features that come handy for our discussion, and in particular the ones related to the characteristics of microservices above.

In Jolie each program is a microservice, and its description is composed by a behaviour, and some deployment information, concerning how it can communicate with other microservices. In this sense, distribution is inherent in the language, since each microservice makes its functionalities available at a specific URL, and can be invoked by other microservices. Non-uniform scalability can be easily obtained: new microservice instances can be run, and one can easily redirect requests from a single microservice to a load balancer: targets of microservice invocations are a first-class object in Jolie, hence they can be easily and dynamically changed. Primitives for architectural composition such as redirection or aggregation also help in this direction. The support that Jolie provides to these basic aspects, and the fact that it fully supports the microservice paradigm, ensure that also the other relevant properties hold. Indeed, Jolie has no specific language support for containerization or elasticity, and indeed how such a language support can be provided and whether it would be beneficial for the language or not is an active topic of discussion in the Jolie community. However, Jolie microservices can be easily deployed in Docker~\cite{docker} containers or on the cloud, hence what said above in this respect holds for Jolie microservices as well. We close this section with a note on robustness: Jolie provides advanced mechanisms for fault notification between different microservices~\cite{GuidiLMZ09}. These mechanisms allow detailed control on whether faults are propagated from one microservice to the ones interacting with it. Indeed, non propagating them allows one to avoid cascading errors, but careful propagation allows one to restore a correct distributed state for the whole system, while preserving the independence of the single microservices. Indeed, each microservice is responsible for restoring its own state, but distributed coordination allows one to ensure global consistency.



\section{Applications}

Microservice architecture has an ideal application where scalability, minimality and cohesiveness are required. Several companies nowadays are moving their monolithic architectures to microservices to reap benefits of scalability. Netflix is one such example - they were one of the pioneers who moved from monolith to microservices~\cite{netflix}. Now Netflix’ underlying microservice architecture enables them to scale effectively and serve millions of users everyday. Portability was used by Netflix not only to make deployment and relocation easier, but also to automatise the deployment: a deployment tool that knew how to deploy a container, could deploy it no matter what was inside it. Microservice architecture also allowed Netflix to improve robustness and availability by launching a service called “Chaos monkey”~\cite{chaosm} to continuously test for faults within the system. Chaos monkey, as the name suggests, causes chaos inside the system by shutting down various services randomly and observing how the system would adapt to these failures. Despite the fact that Chaos Monkey produces faults on the running system, it still operates within the limited period of time when engineers are able to respond to the possible crash.

Our research group has investigated another application of the architectural style exploiting the flexibility of the Jolie programming language: the emerging area of
smart buildings, with an outlook on IoT and smart cities. Rooms of a building have been equipped with a number of devices and sensors in order to capture the fundamental parameters determining well-being, comfort and livability of humans, such as temperature, humidity, and illumination \cite{Salikhov2016a,Salikhov2016b}. The purpose is to monitor and optimize working conditions and the software infrastructure, tightly connected to the hardware, makes use of Jolie and microservices. 

The system is designed separating the logic into small components. Each service is responsible  for managing one sensor or one specific function. Some services are written in Java for a simpler interaction with devices, and Jolie works as an orchestrator for the entire set of services. There are several advantages in this approach. First of all, \textit{reusability}. The system supports different kinds of sensors, but the central logic of data extraction is unchanged even when sensors are added, removed or substituted. Second, code \textit{readability} since services are simple components with a simple logic and a clear naming convention. The combination of readability and reusability also leads to \textit{reduced bugs}. \textit{Scalability}, \textit{minimality} and \textit{cohesiveness} are necessary due to the need of connecting sensors and actuators, removing them, adding new ones, manage faults and monitor the dynamical nature of the infrastructure, especially when mobile devices and \textit{"things"} are part of the system. The \textit{elasticity} of the context has to be managed partially automatically, partially through human intervention from a central control panel, therefore demanding the need for service orchestration and workflow management.


\section{Microservices and Beyond}

The microservice architecture does not build on vacuum and relates to well-established paradigms such as OO and SOA. In \cite{ms-pause} a comprehensive survey on recent developments of microservice architecture is presented focusing on the \emph{evolutionary} aspects more than the \emph{revolutionary} ones. The presentation there is intended to help the reader in understanding microservices, their origin and their possible future.


Microservices can be built using a wide range of technologies combined into the same system. However, we support the idea that a language-based approach can simplify development. Jolie is the only language we are aware of that is natively supporting the paradigm. Other workflow languages are capable of describing service orchestration, for example WS-BPEL~\cite{WS-BPEL}.  WS-BPEL provides indeed many of the features necessary to describe workflows of services, plus communication aspects (ports, interfaces). Dynamic workflow reconfiguration can be expressed too~\cite{MazzaraADB11}. However, WS-BPEL has been designed for high-level orchestration, while  programming the internal logic of a single micro-service requires fine-grained procedural constructs. 

Our research team has been deeply involved in the microservice community and actively contributed to its broader adoption. As an open source project, Jolie has already built a community of developers worldwide - both in industry and in academia - taking care of the development, continuously improving its usability, and therefore broadening the adoption. Recent developments and contributions from our team are: extension of the type system \cite{Safina2016}, development of static type checking \cite{Tchitchigin16}, addition of more iterative control structures to support programming, and inline automatic documentation \cite{Bandura2016}. These works geared up the development environment, and started the process of transforming it into a full suite that makes the entire concept attractive to developers and marketable to companies.

The future is certainly not challenge-free. Security of the  paradigm is an issue almost fully untouched \cite{ms-pause}. Commercial-level quality packages for development are still far to come, despite the acceleration in the interest regarding the matter. Fully-verified software is an open problem the same way it is for more traditional development models. A main open problem is how microservices may integrate with the two main emerging platforms, which will likely dominate the near future: the cloud and the Internet of Things. While microservices seem ideal to run on the cloud, thanks to their properties of portability and elasticity, running on the Internet of Things still present some difficulties. In particular, many things have low computational capabilities and present higher risks from a security point of view, since they are easier to compromise. As an example of this second point just consider that botnets such as Mirai~\cite{mirai} are composed by things (routers, IP cameras, digital video recorders, \dots) which normally have very low protection (e.g., passwords fixed by the manufacturer and never changed). Hence integration of microservices and the Internet of Things would make the need for specific security solutions even more urgent.



\bibliographystyle{plain}
\bibliography{bibliography}   

\begin{thebibliography}{10}

\bibitem{Bandura2016}
A.~Bandura, N.~Kurilenko, M.~Mazzara, V.~Rivera, L.~Safina, and A.~Tchitchigin.
\newblock Jolie community on the rise.
\newblock In {\em SOCA}, 2016.

\bibitem{Bondi2000}
A.~B. Bondi.
\newblock Characteristics of scalability and their impact on performance.
\newblock In {\em WOSP}, page 195–203, 2000.

\bibitem{chaosm}
A.~Tseitlin C.~Bennett.
\newblock {Chaos Monkey Released Into The Wild}.
\newblock
  \url{http://techblog.netflix.com/2012/07/chaos-monkey-released-into-wild.html},
  (2012).

\bibitem{AlmeidaALGM04}
E.~Santana de~Almeida, A.~Alvaro, D.~Lucr{\'{e}}dio, V.~Cardoso Garcia, and
  S.~Romero de~Lemos~Meira.
\newblock Rise project: Towards a robust framework for software reuse.
\newblock In {\em IRI}, pages 48--53, 2004.

\bibitem{ms-pause}
N.~Dragoni, M.~Mazzara, S.~Giallorenzo, F.~Montesi, A.~Lluch Lafuente,
  R.~Mustafin, and L.~Safina.
\newblock Microservices: yesterday, today, and tomorrow.
\newblock In {\em Present and Ulterior Software Engineering}. Springer Berlin
  Heidelberg, 2017.

\bibitem{netflix}
M.~McGarr E.~Bukoski, B.~Moyles.
\newblock {How We Build Code at Netflix}.
\newblock
  \url{http://techblog.netflix.com/2016/03/how-we-build-code-at-netflix.html},
  (2016).

\bibitem{evans2004domain}
E.~Evans.
\newblock {\em Domain-driven design: tackling complexity in the heart of
  software}.
\newblock Addison-Wesley Professional, 2004.

\bibitem{GuidiLMZ09}
C.~Guidi, I.~Lanese, F.~Montesi, and G.~Zavattaro.
\newblock Dynamic error handling in service oriented applications.
\newblock {\em Fundam. Inform.}, 95(1):73--102, 2009.

\bibitem{MazzaraADB11}
M.~Mazzara, F.~Abouzaid, N.~Dragoni, and A.~Bhattacharyya.
\newblock Design, modelling and analysis of a workflow reconfiguration.
\newblock In {\em International Workshop on Petri Nets and Software
  Engineering}, pages 10--24, 2011.

\bibitem{docker}
D.~Merkel.
\newblock Docker: lightweight linux containers for consistent development and
  deployment.
\newblock {\em Linux Journal}, 2014(239):2, 2014.

\bibitem{mirai}
Mirai botnet - wikipedia.
\newblock \url{https://en.wikipedia.org/wiki/Mirai\_(malware)}.

\bibitem{MGZ14}
F.~Montesi, C.~Guidi, and G.~Zavattaro.
\newblock {Service-Oriented Programming with Jolie}.
\newblock In {\em Web Services Foundations}, pages 81--107. Springer, 2014.

\bibitem{Neuman94}
B.~Clifford Neuman.
\newblock Scale in distributed systems.
\newblock In {\em Readings in Distributed Computing Systems}, page 463–489.
  IEEE Computer Society Press, 1994.

\bibitem{newman2015building}
S.~Newman.
\newblock {\em Building Microservices}.
\newblock O'Reilly Media, Inc., 2015.

\bibitem{WS-BPEL}
OASIS.
\newblock {\em Web Services Business Process Execution Language Version 2.0},
  2007.
\newblock \url{http://docs.oasis-open.org/wsbpel/2.0/OS/wsbpel-v2.0-OS.pdf}.

\bibitem{Safina2016}
L.~Safina, M.~Mazzara, F.~Montesi, and V.~Rivera.
\newblock Data-driven workflows for microservices (genericity in jolie).
\newblock In {\em AINA}, 2016.

\bibitem{Salikhov2016b}
D.~Salikhov, K.~Khanda, K.~Gusmanov, M.~Mazzara, and N.~Mavridis.
\newblock Jolie good buildings: Internet of things for smart building
  infrastructure supporting concurrent apps utilizing distributed
  microservices.
\newblock In {\em CCIT}, pages 48--53, 2016.

\bibitem{Salikhov2016a}
D.~Salikhov, K.~Khanda, K.~Gusmanov, M.~Mazzara, and N.~Mavridis.
\newblock Microservice-based iot for smart buildings.
\newblock In {\em WAINA}, 2017.

\bibitem{Tchitchigin16}
A.~Tchitchigin, L.~Safina, M.~Mazzara, M.~Elwakil, F.~Montesi, and V.~Rivera.
\newblock Refinement types in jolie.
\newblock In {\em Spring/Summer Young Researchers Colloquium on Software
  Engineering, SYRCoSE}, 2016.

\end{thebibliography}
\end{document}